\documentclass[12pt]{article}
\begin{document}
\setcounter{page}{1}
\def\theequation{\arabic{section}.\arabic{equation}}
\def\theequation{\thesection.\arabic{equation}}
\setcounter{section}{0}

\title{On polarization of strange baryons in reactions\\ p + p $\to$ p +
$\Lambda^0$ + K$^+$ and p + p $\to$ p + $\Sigma^0$ + K$^+$\\ near
thresholds}

\author{A. Ya. Berdnikov, Ya. A. Berdnikov~\thanks{E--mail:
berdnikov@twonet.stu.neva.ru} , A. N. Ivanov, V. A. Ivanova, \\
V. F. Kosmach ,\\ M. D. Scadron~\thanks{E--mail:
scadron@physics.arizona.edu, Physics Department, University of
Arizona, Tucson, Arizona 85721, USA} , and N. I. Troitskaya}

\date{\today}

\maketitle

\begin{center}
{\it Department of Nuclear Physics, State Technical University of
St. Petersburg, 195251 St. Petersburg, Russian Federation}
\end{center}

\begin{center}
\begin{abstract}
Polarization properties of strange baryons produced in pp reactions, p
+ p $\to$ p + $\Lambda^0$ + K$^+$ and p + p $\to$ p + $\Sigma^0$ +
K$^+$, near thresholds of the final states p$\Lambda^0$K$^+$ and
p$\Sigma^0$K$^+$ are analysed relative to polarizations of colliding
protons. The cross sections for pp reactions are calculated within the
effective Lagrangian approach accounting for strong pp rescattering in
the initial state of colliding protons with a dominant contribution of
the one--pion exchange and strong final--state interaction of daughter
hadrons (Eur. Phys. J. {\bf A} 9, 425 (2000)).
\end{abstract}
\end{center}

\newpage

\section{Introduction}
\setcounter{equation}{0}

\hspace{0.2in} Recently [1] we have considered a production of
strangeness in pp reactions, p + p $\to$ p + Y + K$^+$, where ${\rm Y}
= \Lambda^0$ or $\Sigma^0$, near thresholds of daughter hadrons. We
have derived the effective Lagrangian
\begin{eqnarray}\label{label1.1}
{\cal L}^{\rm pp\to pY K^+}(x) &=& i\,\frac{1}{4}\,C_{\rm pY
K^+}\,\varphi^{\dagger}_{\rm K^+}(x)\,
\{[\bar{p}(x)\gamma^5Y^c(x)][\bar{p^c}(x)p(x)]\nonumber\\ &+&
[\bar{p}(x)Y^c(x)][\bar{p^c}(x)\gamma^5p(x)] +
[\bar{p}(x)\gamma^{\mu}Y^c(x)][\bar{p^c}(x)\gamma_{\mu}\gamma^5p(x)]\},
\end{eqnarray}  
describing the effective vertex of the transition p + p $\to$ Y +
$K^+$ + p, where $p(x)$, ${\rm Y}(x)$ and $\varphi_{\rm K^+}(x)$ are
the interpolating operators of the proton, hyperon and K$^+$--meson
fields, the index ($c$) stands for a charge conjugation. The first and
the last two terms in the Lagrangian Eq.(\ref{label1.1}) describe the
pY--pair coupled in the spin singlet, $S=0$, and spin triplet state,
$S=1$, respectively. The coupling constant $C_{\rm pY K^+}$ has been
calculated in Ref.[1] and reads
\begin{eqnarray}\label{label1.2}
C_{\rm pY K^+} = \frac{\displaystyle g_{pY K^+} g^2_{\pi
NN}}{\displaystyle M_{\rm p} + M_{\rm Y} + M_{\rm K^+}
}\,\frac{1}{\displaystyle M^2_{\pi} + 2M_{\rm p}(E_{\vec{p}} - M_{\rm
p})},
\end{eqnarray}  
where $E_{\vec{p}} = \sqrt{ \vec{p}{\,^2} + M^2_{\rm p}}|_{\textstyle
\vec{p} =\vec{p}_0}$ and $p_0 = \sqrt{(M_{\rm Y} + M_{\rm K^+} -
M_{\rm p})(M_{\rm Y} + M_{\rm K^+} + 3 M_{\rm p})}/2$ is the relative
3--momentum of the colliding protons near threshold, $g_{pY K^+}$ and
$g_{\pi NN}$ are the pseudoscalar meson--baryon--baryon coupling
constants [1,2]. Then, $M_{\rm p}$, $M_{\rm Y}$ and $ M_{\rm K^+}$ are
masses of the proton, the hyperon and the K$^+$--meson. The appearance
of the $\pi$--meson mass $M_{\pi}$ testifies the calculation of the
effective coupling constant $C_{\rm pY K^+}$ in the one--pion exchange
approximation. As has been shown in Ref.[1] the accuracy of this
approximation makes up a few percent.

According to relativistically covariant partial--wave analysis
developed by Anisovich {\it et al.} [3] the spin triplet state, $S =
1$, of the pY--pair is splitted into the ${^3}{\rm P}_0$ state,
described by the the second term in (\ref{label1.1}), and the
${^3}{\rm S}_1$ and ${^3}{\rm D}_1$ states mixed in the third term of
(\ref{label1.1}).

In Ref.[1] the cross sections for the reactions p + p $\to$ p +
$\Lambda$ + K$^+$ and p + p $\to$ p + $\Sigma^0$ + K$^+$, calculated
for unpolarized particles, fit experimental data [3--6] with accuracy
better than 11$\%$ for excess of energy $\varepsilon$, defined by
$\varepsilon = \sqrt{s} - M_{\rm p} - M_{\rm Y} - M_{\rm K^+}$ [1],
ranging values from the region $0.68 \,{\rm MeV} \le \varepsilon \le
138\,{\rm MeV}$ [1].

In this paper we calculate the cross sections for the reactions p + p
$\to$ p + Y + K$^+$, where ${\rm Y} = \Lambda^0$ or $\Sigma^0$, near
thresholds in dependence on polarizations of baryons. We analyse
the contributions of the pY--pair produced in the ${^1}{\rm S}_0$,
${^3}{\rm P}_0$, ${^3}{\rm S}_1$ and ${^3}{\rm D}_1$ states. We show
that the pY--pair can be created only in the spin singlet ${^1}{\rm
S}_0$ and spin triplet ${^3}{\rm S}_1$ states. Therewith a production
of a polarized strange baryon relative to polarizations of colliding
protons comes about only for the spin triplet ${^3}{\rm S}_1$ state of
the pY-pair.

The paper is organized as follows. In section 2 the projection
operators introduced by Anisovich {\it et al.} [3] for the projection
of the wave function of a nucleon--nucleon pair onto the ${^3}{\rm
S}_1$ and ${^3}{\rm D}_1$ states are generalized for the case of
non--equal masses of coupled baryons.  In section 3 we calculate the
amplitude of the reaction p + p $\to$ p + Y + K$^+$. We show that near
threshold the pY--pair can be produced only in the spin singlet
${^1}{\rm S}_0$ and spin triplet ${^3}{\rm S}_1$ states. This
corresponds the colliding protons coupled in the ${^3}{\rm P}_0$ and
${^3}{\rm P}_1$ states, respectively. In section 4 we calculate the
cross sections for pp reactions p + p $\to$ p + $\Lambda^0$ + K$^+$
and p + p $\to$ p + $\Sigma^0$ + K$^+$ in dependence on polarizations
of colliding protons and strange baryons $\Lambda^0$ and
$\Sigma^0$. In the Conclusion we discuss the obtained results.

\section{Partial--wave decomposition of effective vertex of 
transition {\rm p + p $\to$ p + Y + K$^+$}} 
\setcounter{equation}{0}

\hspace{0.2in} The calculation of the amplitude of the reaction p + p
$\to$ p + Y + K$^+$ we start with the decomposition of the effective
vertex of the transition p + p $\to$ p + Y + K$^+$, described by the
effective Lagrangian (\ref{label1.1}), into the interactions for which
the pY--pair couples to the initial protons and the K$^+$--meson in
the states with certain orbital momenta. For this aim it is convenient
to pass into momentum representation. In momentum representation the
effective vertex described by the effective Lagrangian
(\ref{label1.1}) reads [1]
\begin{eqnarray}\label{label2.1}
\hspace{-0.5in}&&{\cal M}({\rm pp\to pY K^+}) =\nonumber\\ 
\hspace{-0.5in}&&= i\,\frac{1}{2}\,C_{\rm
pY K^+}\, \Big\{\Big[\bar{u}\Big(-\vec{q}_{\rm pY} -
\frac{1}{2}\,\vec{p}_{\rm K},\alpha_{\rm
p}\Big)\gamma^5u^c\Big(\vec{q}_{\rm pY} - \frac{1}{2}\,\vec{p}_{\rm
K},\alpha_{\rm Y}\Big)\Big]
\Big[\bar{u^c}(-\vec{p},\alpha_2)u(\vec{p},\alpha_1)\Big]\nonumber\\ 
\hspace{-0.5in}&&+ \Big[\bar{u}\Big(-\vec{q}_{\rm pY} -
\frac{1}{2}\,\vec{p}_{\rm K},\alpha_{\rm p}\Big)u^c\Big(\vec{q}_{\rm
pY} - \frac{1}{2}\,\vec{p}_{\rm K},\alpha_{\rm Y}\Big)\Big]
\Big[\bar{u^c}(-\vec{p},\alpha_2)\gamma^5u(\vec{p},\alpha_1)\Big]
\nonumber\\
\hspace{-0.5in}&& + \Big[\bar{u}\Big(-\vec{q}_{\rm pY} -
\frac{1}{2}\,\vec{p}_{\rm K},\alpha_{\rm
p}\Big)\gamma^{\mu}u^c\Big(\vec{q}_{\rm pY} -
\frac{1}{2}\,\vec{p}_{\rm K},\alpha_{\rm Y}\Big)\Big]
\Big[\bar{u^c}(-\vec{p},\alpha_2)\gamma_{\mu}\gamma^5
u(\vec{p},\alpha_1)\Big]\Big\},
\end{eqnarray}  
According to classification given by Anisovich {\it et al.} [3] the
first, second and third terms in the r.h.s. of (\ref{label2.1})
describe the contribution of the pY--pair coupled in the ${^1}{\rm
S}_0$, ${^3}{\rm P}_0$ and a mixture of ${^3}{\rm S}_1$ and ${^3}{\rm
P}_1$ states respectively. In the low--energy limit there survive only
the first and the third terms of (\ref{label2.1}). Indeed, the wave
function of the pY--pair in the ${^3}{\rm P}_0$ state is proportional
to a relative 3--momentum of the pY--pair and vanishes in the
low--energy limit.  Therefore, near threshold of the reaction p + p
$\to$ p + Y + K$^+$ the second term in (\ref{label2.1}) can be dropped
out. This gives
\begin{eqnarray}\label{label2.2}
\hspace{-0.5in}&&{\cal M}({\rm pp\to pY K^+}) =\nonumber\\
\hspace{-0.5in}&&= i\,\frac{1}{2}\,C_{\rm
pY K^+}\, \Big\{\Big[\bar{u}\Big(-\vec{q}_{\rm pY} -
\frac{1}{2}\,\vec{p}_{\rm K},\alpha_{\rm
p}\Big)\gamma^5u^c\Big(\vec{q}_{\rm pY} - \frac{1}{2}\,\vec{p}_{\rm
K},\alpha_{\rm Y}\Big)\Big]
\Big[\bar{u^c}(-\vec{p},\alpha_2)u(\vec{p},\alpha_1)\Big]\nonumber\\ 
\hspace{-0.5in}&&+ \Big[\bar{u}\Big(-\vec{q}_{\rm pY} -
\frac{1}{2}\,\vec{p}_{\rm K},\alpha_{\rm
p}\Big)\gamma^{\mu}u^c\Big(\vec{q}_{\rm pY} -
\frac{1}{2}\,\vec{p}_{\rm K},\alpha_{\rm Y}\Big)\Big]
\Big[\bar{u^c}(-\vec{p},\alpha_2)\gamma_{\mu}\gamma^5
u(\vec{p},\alpha_1)\Big]\Big\},
\end{eqnarray}  
For the decomposition of the wave function of the pY--pair in the last
term of (\ref{label2.2}) into the states ${^3}{\rm S}_1$ and ${^3}{\rm
D}_1$ with certain orbital momenta we introduce the notations: $k_{\rm
Y} =(E_{\rm Y},\vec{q}_{\rm pY} - \frac{1}{2}\,\vec{p}_{\rm K})
= (E_{\rm Y},\vec{k}_{\rm Y})$, $k_{\rm p} = (E_{\rm p},
-\vec{q}_{\rm pY} - \frac{1}{2}\,\vec{p}_{\rm K}) = (E_{\rm p},
\vec{k}_{\rm p})$, $P = k_{\rm Y} + k_{\rm p}$, $k =
\frac{1}{2}\,(k_{\rm Y} - k_{\rm p})$ and
\begin{eqnarray}\label{label2.3}
\gamma^{\perp}_{\mu} = \gamma_{\mu} -
\hat{P}\,\frac{P_{\mu}}{P^2}\quad,\quad k^{\perp}_{\mu} = k_{\mu} -
\frac{P\cdot k}{P^2}\,P_{\mu}.
\end{eqnarray}  
The 4--vectors $\gamma^{\perp}_{\mu}$ and $k^{\perp}_{\mu}$ are
orthogonal to $P_{\mu}$: $P\cdot \gamma^{\perp} = P\cdot k^{\perp} =
0$.

The baryon densities describing the pY--pair in the ${^3}{\rm S}_1$
and ${^3}{\rm D}_1$ states are defined by [3] (see also [8])
\begin{eqnarray}\label{label2.4}
\Psi_{\mu}({^3}{\rm S}_1;\alpha_{\rm p},\alpha_{\rm Y}) &=&
[\bar{u}(k_{\rm p},\alpha_{\rm p}) S_{\mu} u^c(k_{\rm
Y},\alpha_{\rm Y})],\nonumber\\ \Psi_{\mu}({^3}{\rm D}_1;\alpha_{\rm
p},\alpha_{\rm Y}) &=& [\bar{u}(k_{\rm p},\alpha_{\rm
p}) D_{\mu} u^c(k_{\rm Y},\alpha_{\rm Y})],
\end{eqnarray}  
where $S_{\mu}$ and $D_{\mu}$ are relativistically covariant operators
of the projection onto the ${^3}{\rm S}_1$ and ${^3}{\rm D}_1$ states,
respectively:
\begin{eqnarray}\label{label2.5}
S_{\mu} &=& \frac{1}{\sqrt{2}}\,\frac{1}{\sqrt{P^2 - (M_{\rm Y} -
M_{\rm p})^2}}\,\Bigg[\gamma^{\perp}_{\mu} + \frac{2 }{M_{\rm Y} +
M_{\rm p} + \sqrt{P^2}}\,k^{\perp}_{\mu}\Bigg],\nonumber\\ D_{\mu}
&=&\frac{2}{\sqrt{P^2 - (M_{\rm Y} - M_{\rm
p})^2}}\Bigg[\frac{1}{4}\,\Bigg(1 - \frac{(M_{\rm Y} + M_{\rm
p})^2}{P^2}\Bigg)\Bigg(1 - \frac{(M_{\rm Y} - M_{\rm
p})^2}{P^2}\Bigg)\gamma^{\perp}_{\mu}\nonumber\\
&&-\frac{1}{P^2}\,\Bigg(1 - \frac{(M_{\rm Y} - M_{\rm
p})^2}{P^2}\Bigg)\Bigg(\sqrt{P^2} + \frac{M_{\rm Y} + M_{\rm
p}}{2}\Bigg)\,k^{\perp}_{\mu}\Bigg].
\end{eqnarray}  
This is the generalization of the projection operators introduced by
Anisovich {\it et al.} (see (C.2--C.3) of Ref.[3]) for non--equal
masses of coupled baryons.

In the center of mass frame of the pY--pair the baryon densities
(\ref{label2.4}) are equal to
\begin{eqnarray}\label{label2.6}
\hspace{-0.5in}\Psi_0({^3}{\rm S}_1;\alpha_{\rm p},\alpha_{\rm Y}) &=&
[\bar{u}(k_{\rm p},\alpha_{\rm p}) S_0 u^c(k_{\rm Y},\alpha_{\rm Y})]
= 0,\nonumber\\ \hspace{-0.5in}\vec{\Psi}({^3}{\rm S}_1;\alpha_{\rm
p},\alpha_{\rm Y}) &=& [\bar{u}(k_{\rm p},\alpha_{\rm p}) \vec{S}
u^c(k_{\rm Y},\alpha_{\rm Y})] =
\frac{1}{\sqrt{2}}\,\varphi^{\dagger}_{\rm p}(\alpha_{\rm
p})\vec{\sigma}\varphi_{\rm Y}(\alpha_{\rm Y}),\nonumber\\
\hspace{-0.5in}\Psi_0({^3}{\rm D}_1;\alpha_{\rm p},\alpha_{\rm Y}) &=&
[\bar{u}(k_{\rm p},\alpha_{\rm p}) D_0 u^c(k_{\rm Y},\alpha_{\rm Y})]
= 0,\nonumber\\ \hspace{-0.5in}\vec{\Psi}({^3}{\rm D}_1;\alpha_{\rm
p},\alpha_{\rm Y}) &=& [\bar{u}(k_{\rm p},\alpha_{\rm p}) \vec{D}
u^c(k_{\rm Y},\alpha_{\rm Y})] =
- v^2\,\varphi^{\dagger}_{\rm p}(\alpha_{\rm
p})\Bigg(\frac{3(\vec{\sigma}\cdot \vec{n})\,\vec{n} -
\vec{\sigma}}{2}\Bigg)\varphi_{\rm Y}(\alpha_{\rm Y}),
\end{eqnarray}  
where $\vec{n} = \vec{p}/|\vec{p}|$ is a unit vector of a relative
momentum $\vec{p}$ and $v$ amounts to
\begin{eqnarray}\label{label2.7}
v = \sqrt{\Bigg(1 - \frac{(M_{\rm Y} - M_{\rm p})^2}{P^2}\Bigg)\Bigg(1
- \frac{(M_{\rm Y} + M_{\rm p})^2}{P^2}\Bigg)}.
\end{eqnarray}  
Hence, the formulas (\ref{label2.6}) demonstrate that the baryon
densities (\ref{label2.4}) describe the pY--pair in the S-- and
D--wave states with a total spin $S = 1$ and a total momentum $J =
1$. The baryon densities (\ref{label2.6}) are normalized by [8]
\begin{eqnarray}\label{label2.8}
\frac{1}{3}\sum_{\alpha_{\rm p} = \pm 1/2}\sum_{\alpha_{\rm Y} = \pm
1/2}\vec{\Psi}^{\dagger}({^3}{\rm S}_1; \alpha_{\rm p}, \alpha_{\rm
Y})\cdot \vec{\Psi}({^3}{\rm S}_1; \alpha_{\rm p}, \alpha_{\rm Y}) &=&
1,\nonumber\\ \frac{1}{3}\sum_{\alpha_{\rm p} = \pm
1/2}\sum_{\alpha_{\rm Y} = \pm 1/2}\vec{\Psi}^{\dagger}({^3}{\rm D}_1;
\alpha_{\rm p}, \alpha_{\rm Y})\cdot\vec{\Psi}({^3}{\rm D}_1;
\alpha_{\rm p}, \alpha_{\rm Y}) &=& v^{\,4},
\end{eqnarray}
where $v$ is given by (\ref{label2.7}).  The factor 3 in the
denominator of the l.h.s. of Eq.(\ref{label2.8}) describes the number
of the states of the pY--pair with a total momentum $J = 1$, $2J + 1 =
3$.

For the analysis of nuclear reactions it is convenient to remind that
the normalization (\ref{label2.8}) corresponds to the normalization in
the phase volume of the pY--pair [3] (see also [8]):
\begin{eqnarray}\label{label2.9}
\hspace{-0.3in}&&\frac{1}{3}\int {\rm tr}\{L_{\mu}(\hat{k}_{\rm p} +
M_{\rm p})L^{\mu}(-\hat{k}_{\rm Y} + M_{\rm
Y})\}\,(2\pi)^4\delta^{(4)}(P - k_{\rm p} - k_{\rm
Y})\,\frac{d^3k_{\rm p}}{(2\pi)^3 2E_{\rm p}}\, \frac{d^3k_{\rm
Y}}{(2\pi)^3 2E_{\rm Y}} = \nonumber\\
\hspace{-0.3in}&&=\left\{\begin{array}{r@{\quad,\quad}l}
{\displaystyle \frac{1}{8\pi}\Bigg(1 - \frac{(M_{\rm Y} - M_{\rm
p})^2}{P^2}\Bigg)^{1/2}\Bigg(1 - \frac{(M_{\rm Y} + M_{\rm
p})^2}{P^2}\Bigg)^{1/2}} & L_{\mu} = S_{\mu},\\
{\displaystyle \frac{1}{8\pi}\Bigg(1 - \frac{(M_{\rm Y} - M_{\rm
p})^2}{P^2}\Bigg)^{5/2}\Bigg(1 - \frac{(M_{\rm Y} + M_{\rm
p})^2}{P^2}\Bigg)^{5/2}} & L_{\mu} = D_{\mu}.
\end{array}\right.
\end{eqnarray}
Solving equations (\ref{label2.5}) with respect to
$\gamma^{\perp}_{\mu}$ we express $\gamma^{\perp}_{\mu}$ in terms of
the projection operators $S_{\mu}$ and $D_{\mu}$
\begin{eqnarray}\label{label2.10}
\gamma^{\perp}_{\mu} &=& \frac{2\sqrt{2}}{3}\,\Bigg(\sqrt{P^2} +
\frac{M_{\rm Y} + M_{\rm p}}{2}\Bigg)\sqrt{1 - \frac{(M_{\rm Y} - M_{\rm
p})^2}{P^2}}\,S_{\mu}\nonumber\\
&&+ \,\frac{2}{3}\,\frac{(P^2)^{3/2}}{(\sqrt{P^2}
+M_{\rm Y} + M_{\rm p})\sqrt{P^2 - (M_{\rm Y} - M_{\rm
p})^2}}\,D_{\mu}.
\end{eqnarray}
In the limit of equal masses $ M_{\rm Y}= M_{\rm p} = M_{\rm N}$ the
r.h.s. of (\ref{label2.10}) reduces itself to the form of Eq.(2.16) of
Ref.[8].

Near threshold of the reaction p + p $\to$ p + Y + K$^+$ we can define
$\sqrt{P^2}$ in terms of an excess of energy $\varepsilon$:
$\sqrt{P^2} = \varepsilon + M_{\rm Y} + M_{\rm p}$. Hence, near
threshold of the reaction p + p $\to$ p + Y + K$^+$ the Dirac matrix
$\gamma^{\perp}_{\mu}$ expanded into the projection operators
$S_{\mu}$ and $D_{\mu}$ can be approximated by
\begin{eqnarray}\label{label2.11}
\gamma^{\perp}_{\mu} = 2\sqrt{2}\,\sqrt{M_{\rm Y} M_{\rm p}}\,S_{\mu}
+ \frac{(M_{\rm Y} +  M_{\rm p})^2}{6\sqrt{M_{\rm Y} M_{\rm
p}}}\,D_{\mu}.
\end{eqnarray}
Since in the low--energy limit the parameter $v$ is of order
$O(\varepsilon)$, $v \sim \varepsilon/(M_{\rm Y} + M_{\rm p})$, the
contribution of the ${^3}{\rm D}_1$ states can be neglected near
threshold of the reaction p + p $\to$ p + Y + K$^+$.

Substituting (\ref{label2.11}) in (\ref{label2.2}) and keeping only
leading terms in the low--energy limit we arrive at the effective
vertex of the transition p + p $\to$ p + Y + K$^+$ given by
\begin{eqnarray}\label{label2.12}
\hspace{-0.3in}&&{\cal M}({\rm pp\to pY K^+}) = i\,\frac{1}{2}\,C_{\rm
pY K^+}\,\{[\bar{u}(\vec{k}_{\rm p},\alpha_{\rm
p})\gamma^5u^c(\vec{k}_{\rm Y},\alpha_{\rm
Y})][\bar{u^c}(-\vec{p},\alpha_2) u(\vec{p},\alpha_1)]\nonumber\\
\hspace{-0.3in}&& - 2\sqrt{2}\,\sqrt{M_{\rm Y} M_{\rm
p}}[\bar{u}(\vec{k}_{\rm p},\alpha_{\rm
p}\Big)\,\vec{S}\,u^c(\vec{k}_{\rm Y},\alpha_{\rm Y})]\cdot
[\bar{u^c}(-\vec{p},\alpha_2)\vec{\gamma}\,\gamma^5
u(\vec{p},\alpha_1)]\}.
\end{eqnarray}  
We have taken into account the fact that near threshold, when we are
able to neglect a 3--momentum of the K$^+$--meson, the pY--pair is
practically in the center of mass frame. This implies that only
spatial components of the projection operator $S_{\mu}$ are material.

The effective vertex (\ref{label2.12}) evidences that near threshold
of the reaction p + p $\to$ p + Y + K$^+$ the pY--pair can be produced
only in the spin singlet ${^1}{\rm S}_0$ and spin triplet ${^3}{\rm
S}_1$ states. This corresponds to colliding protons coupled in the
${^3}{\rm P}_0$ and ${^3}{\rm P}_1$ states, respectively.

\section{Amplitude of reaction  {\rm p + p $\to$ p + Y + K$^+$}} 
\setcounter{equation}{0}

\hspace{0.2in} For the calculation of the amplitude of the reaction p
+ p $\to$ p + Y + K$^+$ we would follow Ref.[1] and take into account
strong pp interaction in the initial state, i.e. pp rescattering p + p
$\to$ p + p. As has been shown in Ref.[1] the effective pp interaction
responsible for the transition p + p $\to$ p + p can be represented in
the local form
\begin{eqnarray}\label{label3.1}
{\cal L}^{\rm pp\to pp}(x) &=& \frac{1}{8}\,C_{\rm
pp}\,\Big\{[\bar{p}(x)p^c(x)][\bar{p^c}(x)p(x)] + [\bar{p}(x)\gamma^5
p^c(x)][\bar{p^c}(x)\gamma^5 p(x)]\nonumber\\ &&+
[\bar{p}(x)\gamma_{\mu}\gamma^5p^c(x)][\bar{p^c}(x)\gamma^{\mu}
\gamma^5p(x)]\Big\},
\end{eqnarray}  
where the coupling constant $C_{\rm pp}$ is equal to [1]
\begin{eqnarray}\label{label3.2}
C_{\rm pp} = \frac{g^2_{\rm \pi NN}}{4\vec{p}^{\,2}}\,{\ell n}\Bigg(1
+ \frac{4\vec{p}^{\,2}}{M^2_{\pi}}\Bigg).
\end{eqnarray}  
By summing up  infinite series of one--proton loop diagrams the
vertices of which are defined by the effective interaction
Eq.(\ref{label3.1}) we arrive at the expressions [1,9]
\begin{eqnarray}\label{label3.3}
&&[\bar{u^c}(-\vec{p},\alpha_2)u(\vec{p},\alpha_1)] \to
\frac{[\bar{u^c}(-\vec{p},\alpha_2)u(\vec{p},\alpha_1)]}{\displaystyle
1 + \frac{C_{\rm pp}}{64\pi^2}\int\frac{d^4k}{\pi^2 i}\,{\rm
tr}\Bigg\{\frac{1}{M_{\rm p} - \hat{k}}\frac{1}{M_{\rm p} - \hat{k} -
\hat{Q}}\Bigg\}},\nonumber\\
&&[\bar{u^c}(-\vec{p},\alpha_2)\gamma^i\gamma^5 u(\vec{p},\alpha_1)]
\to (D^{-1}_{\rm pp}(Q))^{ij}
[\bar{u^c}(-\vec{p},\alpha_2))\gamma_j\gamma^5
u(\vec{p},\alpha_1)],\nonumber\\ &&D^{ij}_{\rm pp}(Q) = g^{ji} +
\frac{C_{\rm pp}}{64\pi^2}\int\frac{d^4k}{\pi^2 i}\,{\rm
tr}\Bigg\{\gamma^i\gamma^5\frac{1}{M_{\rm p} -
\hat{k}}\gamma^j\gamma^5\frac{1}{M_{\rm p} - \hat{k} -
\hat{Q}}\Bigg\},
\end{eqnarray}  
where Latin indices run over $i=1,2,3$ and $Q =
(2\,\sqrt{\vec{p}^{\,2} + M^2_{\rm p}},\vec{0}\,)$.

After the evaluation of momentum integrals and renormalization of wave
functions of the protons [1,9] we obtain the contributions of strong
pp interaction, pp rescattering, in the initial state of the reaction
p + p $\to$ p + Y + K$^+$:
\begin{eqnarray}\label{label3.4}
\hspace{-0.3in}&&[\bar{u^c}(-\vec{p},\alpha_2)u(\vec{p},\alpha_1)] \to
\frac{[\bar{u^c}(-\vec{p},\alpha_2)u(\vec{p},\alpha_1)]}{\displaystyle
1 + \frac{C_{\rm pp}(\vec{p}^{\;2},\Lambda)}{8\pi^2}\,
\frac{|\vec{p}\,|^3}{E_{\vec{p}}}\left[{\ell n}\left(\frac{E_{\vec{p}}
+ |\vec{p}\,|}{E_{\vec{p}} - |\vec{p}\,|}\right) +\pi\,i\,\right]} =
\nonumber\\\hspace{-0.3in}&&\hspace{1in}=
[\bar{u^c}(-\vec{p},\alpha_2)u(\vec{p},\alpha_1)]\,f^{\rm pYK^+}_{\rm
pp}({^3}{\rm P}_0;|\vec{p}\,|)\,e^{\textstyle i\delta^{\rm pYK^+}_{\rm
pp}({^3}{\rm P}_0;|\vec{p}\,|)},\nonumber\\
\hspace{-0.3in}&&[\bar{u^c}(-\vec{p},\alpha_2) \vec{\gamma}\,\gamma^5
u(\vec{p},\alpha_1)] \to
\frac{[\bar{u^c}(-\vec{p},\alpha_2))\vec{\gamma}\,\gamma^5
u(\vec{p},\alpha_1)]}{\displaystyle 1 + \frac{C_{\rm
pp}(\vec{p}^{\;2},\Lambda)}{8\pi^2}\,
\frac{|\vec{p}\,|^3}{E_{\vec{p}}}\left[{\ell n}\left(\frac{E_{\vec{p}}
+ |\vec{p}\,|}{E_{\vec{p}} - |\vec{p}\,|}\right) +\pi\,i\,\right]}=
\nonumber\\\hspace{-0.3in}&&\hspace{1in}=
[\bar{u^c}(-\vec{p},\alpha_2)\vec{\gamma}\,\gamma^5u(\vec{p},\alpha_1)]
\,f^{\rm pYK^+}_{\rm pp}({^3}{\rm P}_1;|\vec{p}\,|)\,e^{\textstyle
i\delta^{\rm pYK^+}_{\rm pp}({^3}{\rm P}_1;|\vec{p}\,|)},
\end{eqnarray}  
where $C_{\rm pp}(\vec{p}^{\;2},\Lambda)$ amounts to [1]
\begin{eqnarray}\label{label3.5}
C_{\rm pp}(\vec{p}^{\;2},\Lambda) = \frac{\displaystyle C_{\rm
pp}}{\displaystyle 1 + \frac{C_{\rm
pp}\vec{p}^{\;2}}{4\pi^2}\left[{\ell n}\left(\frac{\Lambda}{M_{\rm p}}
+\sqrt{1 + \frac{\Lambda^2}{M^2_{\rm p}}}\right) -
\frac{\Lambda}{\sqrt{M^2_{\rm p} + \Lambda^2}}\right]}.
\end{eqnarray}  
The appearance of the cut--off $\Lambda$ is caused by non--trivial
$\vec{p}$--dependent logarithmically divergent contributions.  The
cut--off $\Lambda$ restricts from above 3--momenta of virtual proton
fluctuations and is equal to $\Lambda = 1200\,{\rm MeV}$ [1].

In our model the amplitudes of pp rescattering in the 
${^3}{\rm P}_0$ and ${^3}{\rm P}_1$ states are equal near threshold of
the reaction p + p $\to$ p + Y + K$^+$.  Therefore, below we denote
\begin{eqnarray}\label{label3.6}
\hspace{-0.3in}&&f^{\rm pYK^+}_{\rm pp}({^3}{\rm
P}_0;|\vec{p}\,|)\,e^{\textstyle i\delta^{\rm pYK^+}_{\rm pp}({^3}{\rm
P}_0;|\vec{p}\,|)} = f^{\rm pYK^+}_{\rm pp}({^3}{\rm
P}_1;|\vec{p}\,|)\,e^{\textstyle i\delta^{\rm pYK^+}_{\rm pp}({^3}{\rm
P}_1;|\vec{p}\,|)} =\nonumber\\
\hspace{-0.3in}&& = f^{\rm pYK^+}_{\rm pp}(|\vec{p}\,|)\,e^{\textstyle
i\delta^{\rm pYK^+}_{\rm pp}(|\vec{p}\,|)} = \frac{1}{\displaystyle 1
+ \frac{C_{\rm pp}(\vec{p}^{\,2},\Lambda)}{8\pi^2}\,
\frac{|\vec{p}\,|^3}{E_{\vec{p}}}\,\left[{\ell
n}\left(\frac{E_{\vec{p}} + |\vec{p}\,|}{E_{\vec{p}} -
|\vec{p}\,|}\right) +\pi\,i\,\right]}.
\end{eqnarray}  
As has been shown in Ref.[1] in our model the amplitude of strong
low--energy pY interaction in the final state we can be represented in
Watson's form for the final--state interaction [10] in terms of the
scattering length $a_{\rm pY}$ and the effective range $r_{\rm pY}$ of
low--energy elastic pY scattering:
\begin{eqnarray}\label{label3.7}
f^{\rm pY \to pY}(q_{\rm pY}) = \frac{1}{\displaystyle 1 -
\frac{1}{2}\,a_{\rm pY}r_{\rm pY}q^2_{\rm pY} + i\,a_{\rm pY}q_{\rm
pY}} = f_{\rm pY}(q_{\rm pY})\,e^{\textstyle i\delta_{\rm pY}(q_{\rm
pY})}.
\end{eqnarray}  
According to Balewski {\it et al.} [10] for the description of the
final pY interaction in the reaction p + p $\to$ p + Y + K$^+$ we
would use average values for scattering lengths and effective ranges
in the spin singlet ${^1}{\rm S}_0$ and spin triplet ${^3}{\rm S}_1$
states of the pY--pair: $a_{\rm pY} = -2.0\,{\rm fm}$ and $r_{\rm pY}
= 1.0\,{\rm fm}$ [10]. This assumes that the amplitudes of low--energy
elastic pY scattering in the spin singlet ${^1}{\rm S}_0$ and spin
triplet ${^3}{\rm S}_1$ states are equal
\begin{eqnarray}\label{label3.8}
f_{\rm pY}({^1}{\rm S}_0;q_{\rm pY})\,e^{\textstyle i\delta_{\rm
pY}({^1}{\rm S}_0;q_{\rm pY})} = f_{\rm pY}({^3}{\rm S}_1;q_{\rm
pY})\,e^{\textstyle i\delta_{\rm pY}({^3}{\rm S}_1;q_{\rm pY})} =
f_{\rm pY}(q_{\rm pY})\,e^{\textstyle i\delta_{\rm pY}(q_{\rm pY})}.
\end{eqnarray}  
In Ref.[1] we have shown that this assumption agrees well with
experimental data [4--7].

Accounting for the Coulomb repulsion between the daughter proton and
the K$^+$--meson we obtain the total amplitude of the reaction p + p
$\to$ p + Y + K$^+$ near threshold of the final state
\begin{eqnarray}\label{label3.9}
\hspace{-0.3in}&&{\cal M}({\rm pp \to pY K^+}) = \frac{i}{2}\,C_{\rm
pY K^+}\,f^{\rm pYK^+}_{\rm pp}(|\vec{p}\,|)\,f_{\rm pY}(q_{\rm
pY})\,e^{\textstyle i\delta^{\rm pYK^+}_{\rm pp}(|\vec{p}\,|) +
i\delta_{\rm pY}(q_{\rm pY})}\,\nonumber\\
\hspace{-0.3in}&&\times\,\sqrt{ \frac{M_{\rm pK^+}}{q_{\rm pK^+}}\,\frac{2\pi \alpha
}{\displaystyle e^{\textstyle 2\pi \alpha M_{\rm pK^+}/q_{\rm pK^+}} -
1}}\,\{[\bar{u}(\vec{k}_{\rm p},\alpha_{\rm
p})\gamma^5u^c(\vec{k}_{\rm Y},\alpha_{\rm
Y})][\bar{u^c}(-\vec{p},\alpha_2) u(\vec{p},\alpha_1)]\nonumber\\
\hspace{-0.3in}&& - 2\sqrt{2}\,\sqrt{M_{\rm Y} M_{\rm
p}}[\bar{u}(\vec{k}_{\rm p},\alpha_{\rm
p}\Big)\,\vec{S}\,u^c(\vec{k}_{\rm Y},\alpha_{\rm Y})]\cdot
[\bar{u^c}(-\vec{p},\alpha_2)\vec{\gamma}\,\gamma^5
u(\vec{p},\alpha_1)]\},
\end{eqnarray}  
where the factor depending of the fine structure constant $\alpha
=1/137$ takes into account the Coulomb repulsion between the daughter
proton and the K$^+$ meson at low relative 3--momenta $q_{\rm pK^+}$
[10] ( see also [9]), $M_{\rm pK^+} = M_{\rm p}M_{\rm K^+}/(M_{\rm p}
+ M_{\rm K^+})$ is a reduced mass of the pK$^+$ system.

\section{Cross sections for reactions  {\rm p + p $\to$ p + 
$\Lambda^0$ + K$^+$} and {\rm p + p $\to$ p + $\Sigma^0$ + K$^+$} with
polarized baryons} 
\setcounter{equation}{0}

\hspace{0.2in} The calculation of the cross section for the reaction p
+ p $\to$ p + Y + K$^+$ we carry out in dependence on polarizations of
strange baryon and colliding protons [11].  The polarization vectors
of coupled baryons we define as follows [12]
\begin{eqnarray}\label{label4.1}
\zeta^{\mu}_1 &=& \Bigg(+\frac{\vec{p}\cdot\vec{\zeta}_{\,1}}{M_{\rm
p}},\vec{\zeta}_{\,1} +
\frac{\vec{p}\,(\vec{p}\cdot\vec{\zeta}_{\,1})}{M_{\rm p}(E_{\vec{p}} +
M_{\rm p})}\Bigg),\nonumber\\ \zeta^{\mu}_{\,2} &=&
\Bigg(-\frac{\vec{p}\cdot\vec{\zeta}_{\,2}}{M_{\rm
p}},\vec{\zeta}_{\,2} +
\frac{\vec{p}\,(\vec{p}\cdot\vec{\zeta}_{\,2})}{M_{\rm p}(E_{\vec{p}} +
M_{\rm p})}\Bigg),\nonumber\\ \zeta^{\mu}_{\,{\rm Y}}
&=&(0,\vec{\zeta}_{\,{\rm Y}}),
\end{eqnarray}  
where $\vec{\zeta}_{\,i}\,(i=1,2,{\rm Y})$ are polarization vectors of
baryons normalized to unity $\vec{\zeta}^{\;2}_{\,i} = 1$.

Introducing the polarization vectors of baryons in a standard way
\begin{eqnarray}\label{label4.2}
\sum_{\alpha_1=\pm 1/2}u(p_1,\alpha_1)\bar{u}(p_1,\alpha_1) &=&
(\hat{p}_1 + M_{\rm p})\Bigg(\frac{1 +
\gamma^5\hat{\zeta}_1}{2}\Bigg),\nonumber\\ \sum_{\alpha_2=\pm
1/2}u^c(p_2,\alpha_2)\bar{u^c}(p_2,\alpha_2) &=& (\hat{p}_2 - M_{\rm
p})\Bigg(\frac{1 + \gamma^5\hat{\zeta}_2}{2}\Bigg),\nonumber\\
\sum_{\alpha_{\rm Y}=\pm 1/2}u^c(k_{\rm Y},\alpha_{\rm
Y})\bar{u^c}(k_{\rm Y},\alpha_{\rm Y}) &=& (\hat{k}_{\rm Y} - M_{\rm
Y})\Bigg(\frac{1 + \gamma^5\hat{\zeta}_{\rm Y}}{2}\Bigg)
\end{eqnarray}  
we calculate the squared amplitude (\ref{label3.9}),
averaged and summed over the states of colliding protons and final
baryons. The result reads
\begin{eqnarray}\label{label4.3}
\hspace{-0.3in}&&\overline{|{\cal M}({\rm pp \to pY K^+})|^2} =
C^2_{\rm p\Lambda K^+}\,|f^{\rm
pYK^+}_{\rm pp}(|\vec{p}\,|)|^2|f_{\rm pY}(q_{\rm
pY})|^2\,\frac{M_{\rm pK^+}}{q_{\rm pK^+}}\,\frac{2\pi \alpha
}{\displaystyle e^{\textstyle 2\pi \alpha M_{\rm pK^+}/q_{\rm pK^+}} -
1}\nonumber\\
\hspace{-0.3in}&&\times\,4\vec{p}^{\;2}M_{\rm p}M_{\rm Y}\,\Big(1 + \frac{1}{3}\,\vec{\zeta}_1\cdot
\vec{\zeta}_2 + \frac{1}{3}\,\vec{n}\cdot(\vec{\zeta}_1 +
\vec{\zeta}_2)(\vec{n}\cdot \vec{\zeta}_{\rm Y})\Big) =
\overline{|{\cal M}({\rm pp \to pYK^+})|^2}_0\nonumber\\
\hspace{-0.3in}&&\times\,\Big(1 + \frac{1}{3}\,\vec{\zeta}_1\cdot
\vec{\zeta}_2 + \frac{1}{3}\,\vec{n}\cdot(\vec{\zeta}_1 +
\vec{\zeta}_2)(\vec{n}\cdot \vec{\zeta}_{\rm Y})\Big),
\end{eqnarray}  
where $\overline{|{\cal M}({\rm pp \to pY K^+})|^2}_{\!0}$ is a
squared amplitude of the reaction under consideration with unpolarized
particles and $\vec{n} = \vec{p}/| \vec{p}\,|$ is a unit vector
along a relative 3--momentum of colliding protons.

If only one of the colliding protons is polarized, the amplitude
(\ref{label4.3}) reduces to a simpler form
\begin{eqnarray}\label{label4.4}
\overline{|{\cal M}({\rm pp \to pY K^+})|^2} = \overline{|{\cal
M}({\rm pp \to pY K^+})|^2}_0\,\Big(1
+\frac{1}{3}\,(\vec{n}\cdot\vec{\zeta})(\vec{n}\cdot \vec{\zeta}_{\rm
Y})\,\Big),
\end{eqnarray}  
where $\vec{\zeta}$ is a polarization vector of the polarized proton
in the initial state.

Using (\ref{label4.3}), (\ref{label4.4}) and the results obtained in
Ref.[1] we write down the cross sections for the reactions p + p $\to$
p + $\Lambda^0$ + K$^+$ and p + p $\to$ p + $\Sigma^0$ + K$^+$, when
(i) colliding protons and a strange baryon are polarized, $\vec{\rm p}
+ \vec{\rm p} \to {\rm p } + \vec{\rm Y} + {\rm K}^+$, and (ii) there
are polarized only one of the colliding protons and a strange baryon,
$\vec{\rm p} + {\rm p} \to {\rm p } + \vec{\rm Y} + {\rm K}^+$ or
${\rm p} + \vec{\rm p} \to {\rm p } + \vec{\rm Y} + {\rm K}^+$:
\begin{eqnarray}\label{label4.5}
\hspace{-0.5in}\sigma^{\rm \vec{\rm p}\,\vec{\rm p}\to
p\vec{\Lambda}^0 K^+}(\varepsilon) &=& \sigma^{\rm pp\to p\Lambda^0
K^+}(\varepsilon)\,\Big(1 + \frac{1}{3}\,\vec{\zeta}_1\cdot
\vec{\zeta}_2 + \frac{1}{3}\,\vec{n}\cdot(\vec{\zeta}_1 +
\vec{\zeta}_2)(\vec{n}\cdot \vec{\zeta}_{\Lambda^0})\,
\Big),\nonumber\\
\hspace{-0.5in}\sigma^{\rm \vec{\rm p}\,\vec{\rm p}\to p\vec{\Sigma}^0
K^+}(\varepsilon) &=& \sigma^{\rm pp\to p\Sigma^0
K^+}(\varepsilon)\,\Big(1 + \frac{1}{3}\,\vec{\zeta}_1\cdot
\vec{\zeta}_2 + \frac{1}{3}\,\vec{n}\cdot(\vec{\zeta}_1 +
\vec{\zeta}_2)(\vec{n}\cdot
\vec{\zeta}_{\Sigma^0})\Big).
\end{eqnarray}  
For reactions p + $\vec{\rm p}$ $\to$ p + $\vec{\Lambda}^0$ + K$^+$
and p + $\vec{\rm p}$ $\to$ p + $\vec{\Sigma}^0$ + K$^+$ with one
polarized proton in the initial state and a polarized strange baryon
we get
\begin{eqnarray}\label{label4.6}
\hspace{-0.3in}\sigma^{\rm p\,\vec{\rm p}\to p\vec{\Lambda}^0
K^+}(\varepsilon) &=& \sigma^{\rm pp\to p\Lambda^0
K^+}(\varepsilon)\,\Big(1 + \frac{1}{3}\,\,(\vec{n}\cdot
\vec{\zeta})(\vec{n}\cdot
\vec{\zeta}_{\Lambda^0})\Big),\nonumber\\
\hspace{-0.3in}\sigma^{\rm p\,\vec{\rm p}\to p\Sigma^0
K^+}(\varepsilon) &=& \sigma^{\rm pp\to p\Sigma^0
K^+}(\varepsilon)\,\Big(1 +
\frac{1}{3}\,\,(\vec{n}\cdot\vec{\zeta})(\vec{n}\cdot
\vec{\zeta}_{\Sigma^0})\Big).
\end{eqnarray}  
The cross sections for unpolarized baryons $\sigma^{\rm pp\to
p\Lambda^0 K^+}(\varepsilon)$ and $\sigma^{\rm pp\to p\Sigma^0
K^+}(\varepsilon)$ have been tabulated in Ref.[1] for excess of energy
$\varepsilon$ ranging values from the region $0.68\,{\rm MeV} \le
\varepsilon \le 138\,{\rm MeV}$. Theoretical cross sections fit
experimental data with accuracy better than 11$\%$.

\section{Conclusion}
\setcounter{equation}{0}

\hspace{0.2in} We have shown that in our approach [1] to the
description of pp reactions p + p $\to$ p + $\Lambda^0$ + K$^+$ and p
+ p $\to$ p + $\Sigma^0$ + K$^+$ near thresholds of the final states
p$\Lambda$K$^+$ and p$\Sigma^0$K$^+$, based on pp rescattering in the
initial state with a dominant contribution of the one--pion exchange
and strong final--state interaction of daughter hadrons, polarization
properties of strange baryons can be investigated with respect to
polarizations of colliding protons.

Near thresholds of the reactions p + p $\to$ p + $\Lambda^0$ + K$^+$
and p + p $\to$ p + $\Sigma^0$ + K$^+$ we predict production of
p$\Lambda^0$ and p$\Sigma^0$ pairs only in the spin singlet ${^1}{\rm
S}_0$ and spin triplet ${^3}{\rm S}_1$ states. This result has been
obtained by means of relativistically covariant partial--wave analysis
worked out by Anisovich {\it et al.} for nucleon--nucleon scattering
[3]. In order to implement this analysis to reactions p + p $\to$ p +
$\Lambda^0$ + K$^+$ and p + p $\to$ p + $\Sigma^0$ + K$^+$ for the
description of wave functions of p$\Lambda^0$ and p$\Sigma^0$ pairs we
have generalized the projection operators introduced by Anisovich {\it
et al.} for nucleon--nucleon pairs onto the case of baryon--baryon
pairs with non--equal masses of coupled baryon.

In our model production of a polarized strange baryon can come about
only for p$\Lambda^0$ and p$\Sigma^0$ pairs produced in the spin
triplet state ${^3}{\rm S}_1$. The more detailed predictions for
polarization of strange baryons can be obtained from theoretical cross
sections (\ref{label4.5}) and (\ref{label4.6}) in accord specific
experimental conditions of experimental analysis of reactions p + p
$\to$ p + $\Lambda^0$ + K$^+$ and p + p $\to$ p + $\Sigma^0$ + K$^+$.

Now let us compare our results with the model--independent analysis of
polarization of strange baryons in the reaction p + p $\to$ p + Y +
K$^+$ worked out by Rekalo {\it et al.} [11]. According to Ref.[11]
the most general form of the cross section for the reaction $\vec{\rm
p} + \vec{\rm p} \to {\rm p } + {\rm Y} + {\rm K}^+$ with polarized
colliding protons and unpolarized strange baryon should read
\begin{eqnarray}\label{label5.1}
\hspace{-0.5in}\sigma^{\rm \vec{\rm p}\,\vec{\rm p}\to pY
K^+}(\varepsilon) &=& \sigma^{\rm pp\to pY K^+}(\varepsilon)\,\Big(1 +
{\cal A}_1\,\vec{\zeta}_1\cdot \vec{\zeta}_2 + {\cal
A}_2\,(\vec{n}\cdot\vec{\zeta}_1)(\vec{n}\cdot\vec{\zeta}_2)\Big),
\end{eqnarray}  
where ${\cal A}_i\,(i=1,2)$ are real functions obeying the constraint
\begin{eqnarray}\label{label5.2}
3{\cal A}_1 + {\cal A}_2 = 1.
\end{eqnarray}  
When matching the expression (\ref{label5.1}) with ours
(\ref{label4.5}) we find that
\begin{eqnarray}\label{label5.3}
{\cal A}_1 = \frac{1}{3}\quad,\quad {\cal A}_2 = 0.
\end{eqnarray}  
This agrees completely with Rekalo's prediction (\ref{label5.2}).

Unlike our results Rekalo {\it et al.} did not give an explicit
expression of the cross section for the reaction p + p $\to$ p + Y +
K$^+$ with polarized colliding protons and strange baryon. Therefore,
we cannot compare our theoretical cross sections (\ref{label4.5}) and
(\ref{label4.6}) with analogous expressions which could be obtained
within a model--independent approach [11].

However, following general properties of strong interactions and
parity invariance, in particular, Rekalo {\it et al.}  predicted that
the projection of the polarization vector $\vec{\zeta}_{\rm Y}$ of a
strange baryon onto a relative 3--momentum of colliding protons
$\vec{n}\cdot \vec{\zeta}_{\rm Y}$ is proportional to $\vec{n}\cdot
\vec{\zeta}$ [11]
\begin{eqnarray}\label{label5.4}
\vec{n}\cdot \vec{\zeta}_{\rm Y} = (1 - {\cal A}_1)\,\vec{n}\cdot
\vec{\zeta}.
\end{eqnarray}  
In our model, when ${\cal A}_1 = 1/3$, this gives
\begin{eqnarray}\label{label5.5}
\vec{n}\cdot \vec{\zeta}_{\rm Y} =
\frac{2}{3}\,\vec{n}\cdot\vec{\zeta}.
\end{eqnarray}  
This result can be verified experimentally.

The absence in our cross sections (\ref{label4.5}) the terms
$\vec{\zeta}_{\rm Y}\cdot(\vec{\zeta}_1\times \vec{\zeta}_2)$,
$(\vec{n}\cdot\vec{\zeta}_{\rm Y})(\vec{n}\cdot(\vec{\zeta}_1\times
\vec{\zeta}_2))$ and so testifies that in our approach polarization
observables of strange baryons defining the cross section for the
reaction p + p $\to$ p + Y + K$^+$ are even under time reversal,
$T$--even polarization observables [12].  According to Rekalo's
model--independent analysis this assumes the relation between phases
of amplitudes of pp and pY scattering
\begin{eqnarray}\label{label5.6}
\delta^{\rm pYK^+}_{\rm pp}({^3}{\rm P}_0;|\vec{p}\,|) + \delta_{\rm
pY}({^1}{\rm S}_0;q_{\rm pY}) = \delta^{\rm pYK^+}_{\rm pp}({^3}{\rm
P}_1;|\vec{p}\,|) + \delta_{\rm pY}({^3}{\rm S}_1; q_{\rm pY}),
\end{eqnarray}  
where $\delta^{\rm pYK^+}_{\rm pp}({^3}{\rm P}_0;|\vec{p}\,|)$ and
$\delta^{\rm pYK^+}_{\rm pp}({^3}{\rm P}_1;|\vec{p}\,|)$ are the
phases of amplitudes of strong pp rescattering in the ${^3}{\rm P}_0$
and ${^3}{\rm P}_1$ states, respectively, and $\delta_{\rm
pY}({^1}{\rm S}_0;q_{\rm pY})$ and $\delta_{\rm pY}({^3}{\rm
S}_1;q_{\rm pY})$ are the phases of low--energy elastic pY scattering
in the spin singlet ${^1}{\rm S}_0$ and spin triplet ${^3}{\rm S}_1$
states, respectively. Since scattering lengths and effective ranges of
elastic pY scattering have been set equal this implies that
$\delta_{\rm pY}({^1}{\rm S}_0;q_{\rm pY}) = \delta_{\rm pY}({^3}{\rm
S}_1;q_{\rm pY})$. Substituting this relation into (\ref{label5.6}) we
obtain the constraint
\begin{eqnarray}\label{label5.7}
\delta^{\rm pYK^+}_{\rm pp}({^3}{\rm P}_0;|\vec{p}\,|) = \delta^{\rm
pYK^+}_{\rm pp}({^3}{\rm P}_1;|\vec{p}\,|).
\end{eqnarray}  
Hence, any experimental measurement of the cross section for the
reaction p + p $\to$ p + Y + K$^+$ with non--vanishing contributions
of $T$--odd polarization observables like $\vec{\zeta}_{\rm
Y}\cdot(\vec{\zeta}_1\times \vec{\zeta}_2)$ should evidence a
violation of constraints (\ref{label5.6}) and (\ref{label5.7}). The
former might mean that either scattering lengths and effective ranges
of low--energy elastic pY scattering are not really equal for the spin
singlet ${^1}{\rm S}_0$ and spin triplet ${^3}{\rm S}_1$ states or, in
reality, amplitudes of strong pp rescattering in the ${^3}{\rm P}_0$
and ${^3}{\rm P}_1$ states of colliding protons differ themselves near
threshold of the reaction p + p $\to$ p + Y + K$^+$.

\end{document}